\documentclass[12pt,doublecolumn]{iopart}

\usepackage{graphicx}
\usepackage{epsfig}
\usepackage{subfigure}
\usepackage{url}
\usepackage{color}
\begin{document}
\title{Urban Gravity: a Model for Intercity Telecommunication Flows}
\author{Gautier Krings$^{1,2}$, Francesco Calabrese$^{2}$, Carlo Ratti$^{2}$ and Vincent D Blondel$^{1}$}

\address{$^1$ Universit\'e catholique de Louvain (UCL), Department of Applied Mathematics - \\4 Avenue Georges Lemaitre, B-1348 Louvain-la-Neuve, Belgium}
\address{$^2$ SENSEable City Laboratory, Massachusetts Institute of Technology - \\77 Massachusetts Avenue, Cambridge, MA 02139, USA}
\ead{gautier.krings@uclouvain.be}

\begin{abstract} We analyze the anonymous communication patterns of 2.5 million customers of a Belgian mobile phone operator. Grouping customers by billing address, we build a social network of cities, that consists of communications between 571 cities in Belgium. We show that inter-city communication intensity is characterized by a gravity model: the communication intensity between two cities is proportional to the product of their sizes divided by the square of their distance.
\end{abstract}

\maketitle

Recent research has shown that certain characteristics of cities grow in different ways in relation to population size. While some characteristics are directly proportional to a city's population size, instead, other features - such as productivity or energy consumption - are not linear but exhibit superlinear or sublinear dependence to population size \cite{bettencourt2007gis}. Interestingly, some of these features have strong similarities with those found in biological cells - an observation that has led to the creation of a metaphor where cities are seen as living entities \cite{macionis2001cau}.
 Interactions between cities, such as passenger transport flows and phone messages, have also been related to population and distance \cite{zipf1949hba,citeulike:4007493}.
Meanwhile, in socio-economic networks, interactions between entities such as cities or countries have led to models remembering Newton's gravity law, where the sizes of the entities play the role of mass \cite{carrothers1958hrg}. Road and airline networks between cities have also been studied \cite{jung2008gmk,barrat2004acw}, and in the case of road networks, it appears that the strength of interaction also follows a gravity law.

While these results have provided a better understanding of the way cities interact, a finer analysis at human level was until now difficult because of a lack of data. Recently, however, telephone communication data has opened up a new way of analyzing cities at both a fine and aggregate level (whereby as Gottman as already as 1957 noted \cite{gottman1957mun}: ``the density of the flow of telephone calls is a fairly good measure of the relationships binding together the economic interests of the region'').
Several large datasets of email and phone calls have recently become available. By using these as a proxy for social networks, they have enabled the study of  human connections and behaviors \cite{dodds2003ess,onnela2007sat,gonzalez2008uih,onnela2007als,1401963}.
The use of geographical information makes it possible to go one step further in the study of individual and group interactions. For example, Lambiotte \textit{et al.} use a mobile phone dataset to show that the probability for a  call between two people decreases by the square of their distance \cite{lambiotte2008gdm}. However, while the structure of complex networks has already been widely studied \cite{watts1999swd,barabasi2003lec,watts1998cdo,newman2003saf}, to date, contributions  have not yet analyzed large-scale features of social networks where people are aggregated based on their geographical proximity.

In this work, we study anonymized mobile phone communications from a Belgian operator and derive a model of interaction between cities. Grouping customers together by billing address, we create a two-level network, containing both a microscopic network of human-to-human interactions, and a macroscopic network of interactions between cities.\\
The data that we consider consists of the communications made by more than 2.5 million customers of a Belgian mobile phone operator over a period of 6 months in 2006 \cite{lambiotte2008gdm}. Every customer is identified by a surrogate key and to every customer we associate their corresponding billing address zip code.
In order to construct the communication network, we have filtered out calls involving other operators (there are three main operators in Belgium), incoming or outgoing, and we have kept only those transactions in which both the calling and receiving individuals are customers of the mobile phone company.
In order to eliminate ``accidental calls'', we have kept links between two customers $i$ and $j$ only if there are at least six calls in both directions during the 6 months time interval.
The resulting network is composed of 2.5 million nodes and 38 million links. To the link between the customers $i$ and $j$ we associate a communication intensity by computing the total communication time in seconds $l_{ij}$ between $i$ and $j$.\\
In order to analyze the relationship between this social network and geographical positioning, we associate customers to cities based on their billing address zip code. Belgium is a country of approximately 10.5 million inhabitants, with a high population density of 344 inhab./km$^2$.  The Belgian National Institute of Statistics (NIS) \cite{INS} divides this population into 571 cities (cities, towns and villages), whose sizes show an overall lognormal population distribution with approximate parameters $\mu = 4.05$ and $\sigma = 0.37$\footnote{The lognormal distribution of city size is consistent with similar data on US cities \cite{eeckhout2004gsl}.}.\\
\begin{figure}[!h]
	\begin{center}
		\includegraphics[scale = 0.4]{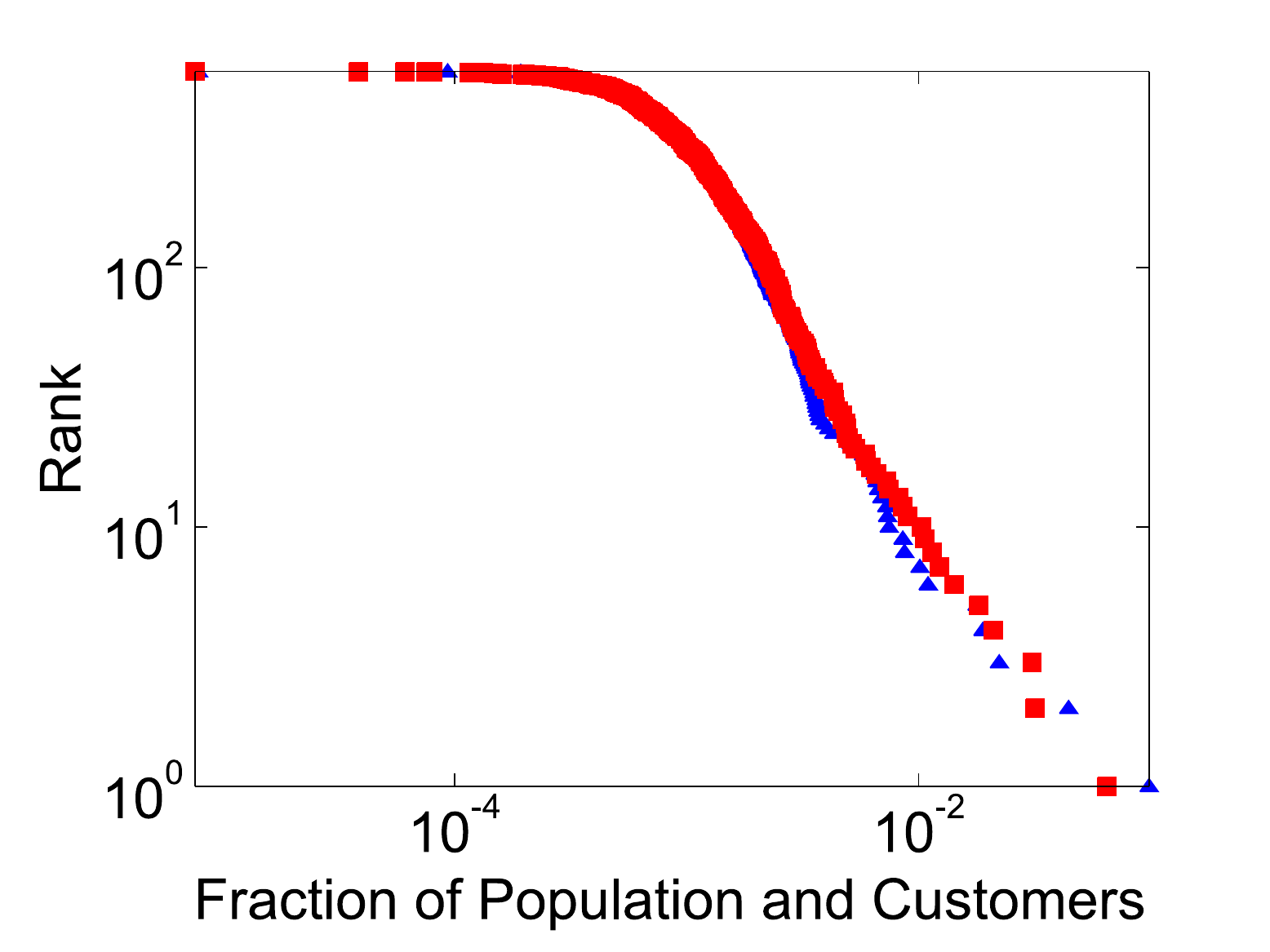}
	\end{center}
	\caption{Ranks of city population sizes (blue triangles) and number of customers (red squares) follow similar distributions.}
\label{fig:popdistr}
\end{figure}
The analyzed communication network provides information for the operator's customers rather than for the entire population. However, the number of customers present in each city follows the same lognormal distribution as the total population and so this suggests that our dataset is not structurally biased by particular user-groups and market shares. This is also confirmed by the ranks of city population sizes that match with those of customers, as shown in Fig. \ref{fig:popdistr}. In the rest of this article, when we use the term \textit{population} of a city, we are refering to the number of customers that have the corresponding ZIP code billing address, even if they are not connected to any other customer in the microscopic network of human-to-human communications defined above. These customers are still taken into account for the population size, since their presence is of interest for the normalization of the communication data.\\
\begin{figure}[!h]
	\begin{center}
		\subfigure[]{\includegraphics[scale = 0.4]{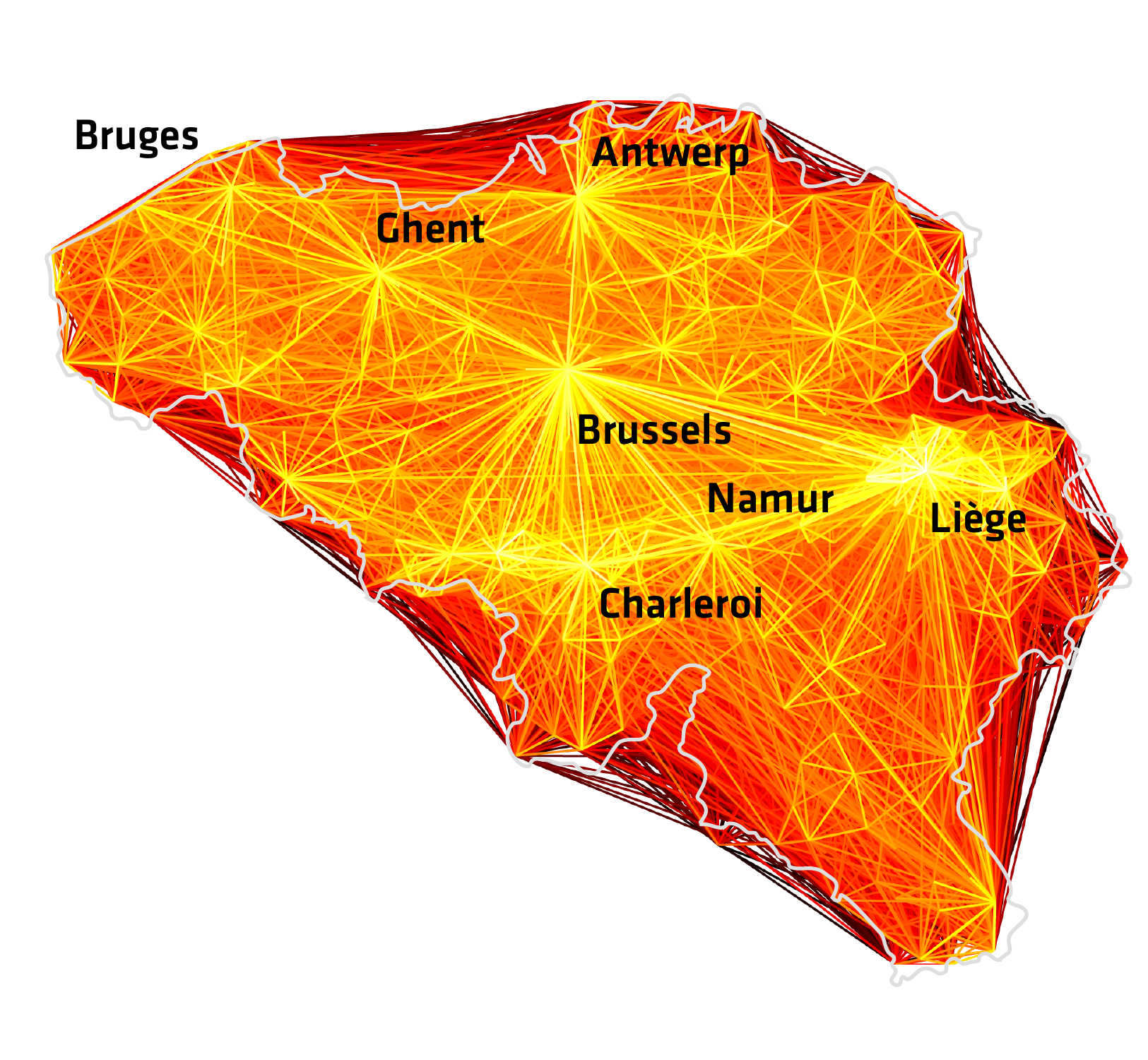}}
		\subfigure[]{\includegraphics[scale = 0.5]{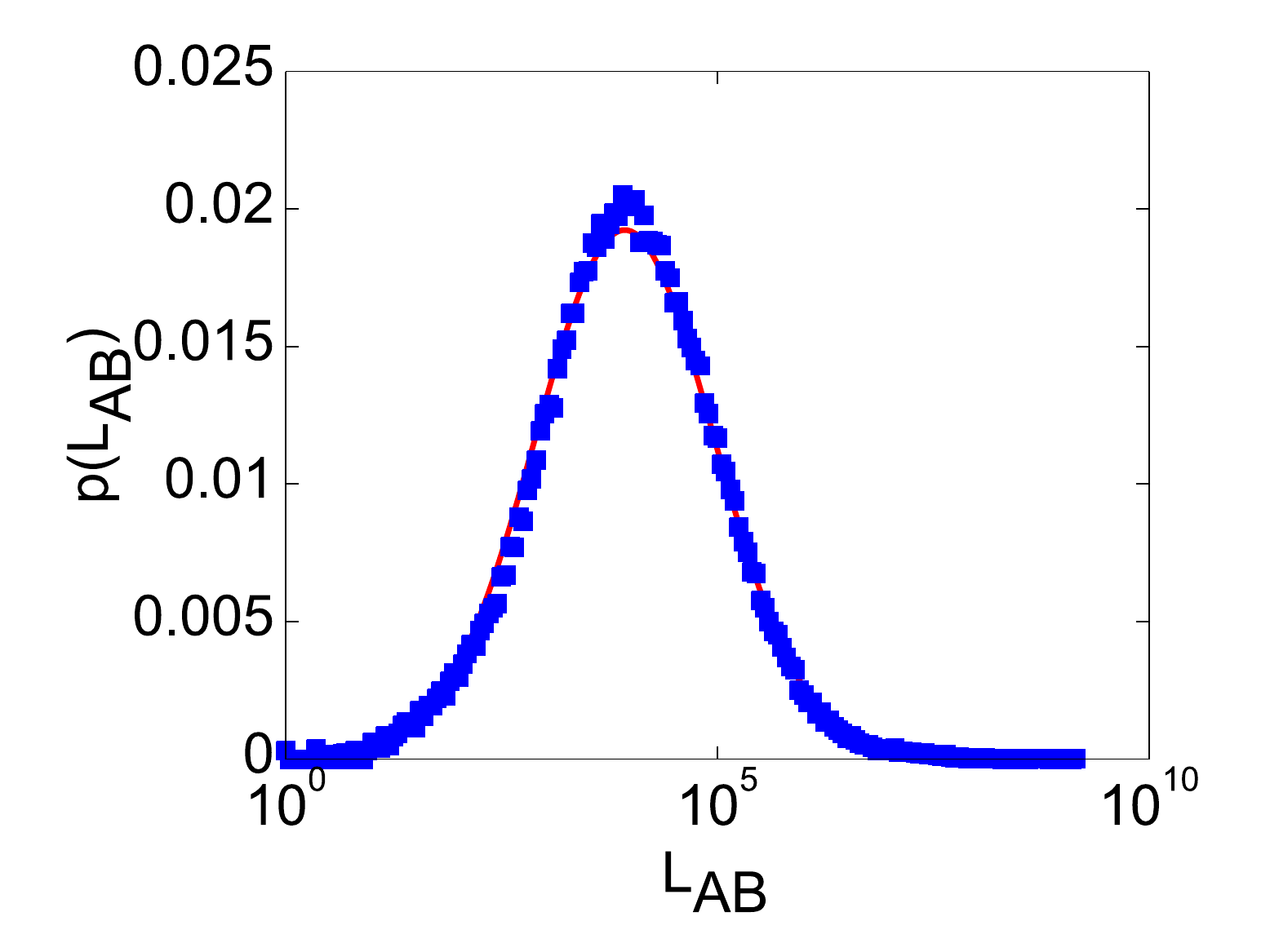}}
	\end{center}
	\caption{(a) Illustration of the macroscopic communication network (only the top 30\% of the links (having the strongest intensity) are represented). Colors indicate the intensity of communication between the cities: bright colors indicate a strong intensity. (b) Intensity distribution of the macroscopic network, self edges are not considered. The red curve shows the lognormal best fit, with parameters $\mu = 3.93$ and $\sigma = 1.03$}
\label{fig:illustration}
\end{figure}
By aggregating the individual communications at a city level, we obtain a network of 571 cities in Belgium. We define the intensity of interaction between the cities $A$ and $B$ by (Fig. \ref{fig:illustration} (a)):
 $$L_{AB} = \sum_{i\in A,\phantom{.}j\in B} l_{ij}.$$
The distribution of intensity is narrow and it appears to follow a lognormal distribution (Fig. \ref{fig:illustration} (b)).  A similar lognormal distribution is found for the degree distribution. This lognormal intensity distribution  is sharply different from what is typically observed in social networks but is consistent with observations in other macroscopic networks, such as the intensity of trade between countries, obtained by aggregating the individual trades made by agents \cite{bhattacharya2008itn}.

\begin{figure}[!h]
	\begin{center}
		\subfigure[]{\includegraphics[scale=0.25]{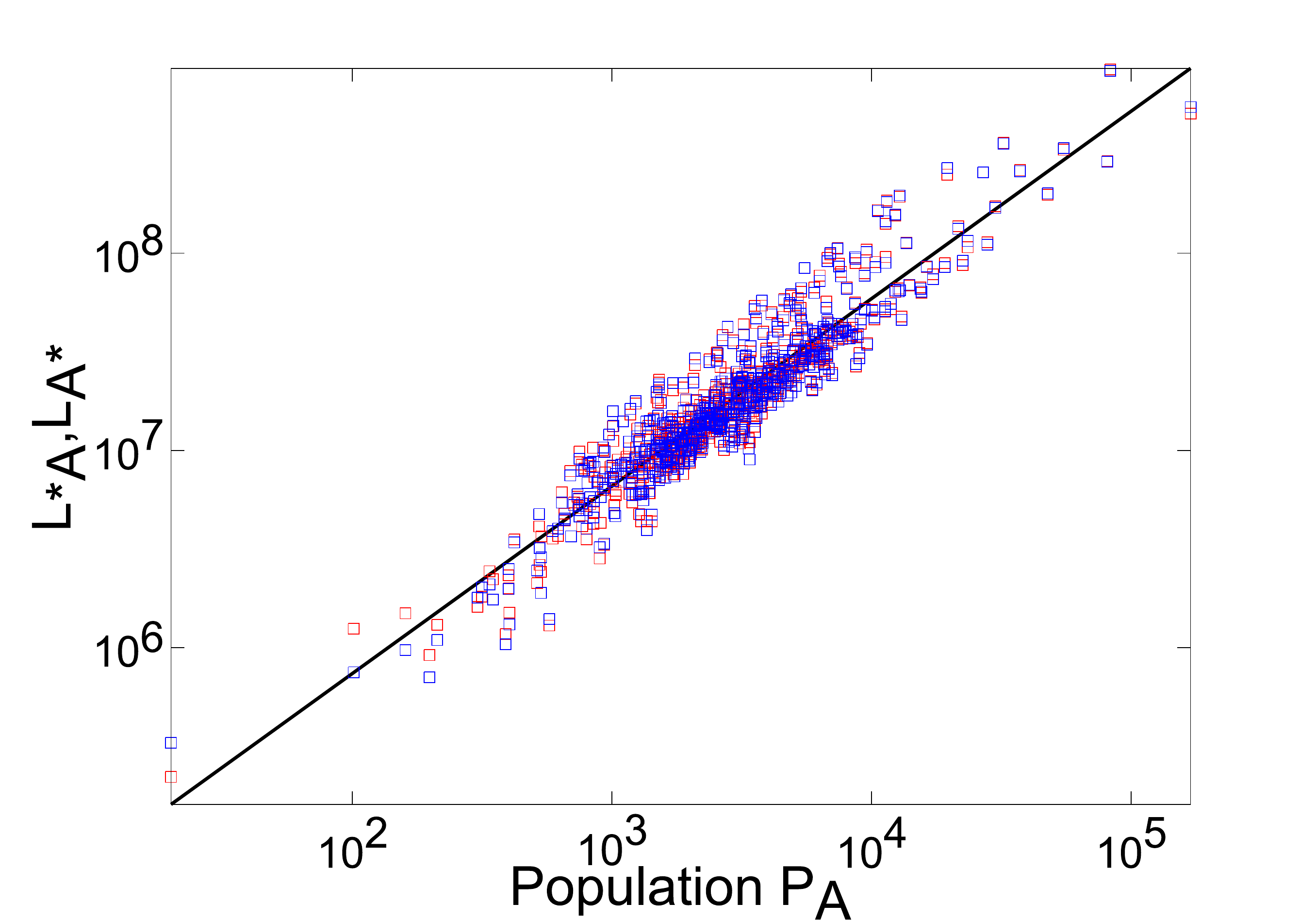}}
		\subfigure[]{\includegraphics[scale=0.4]{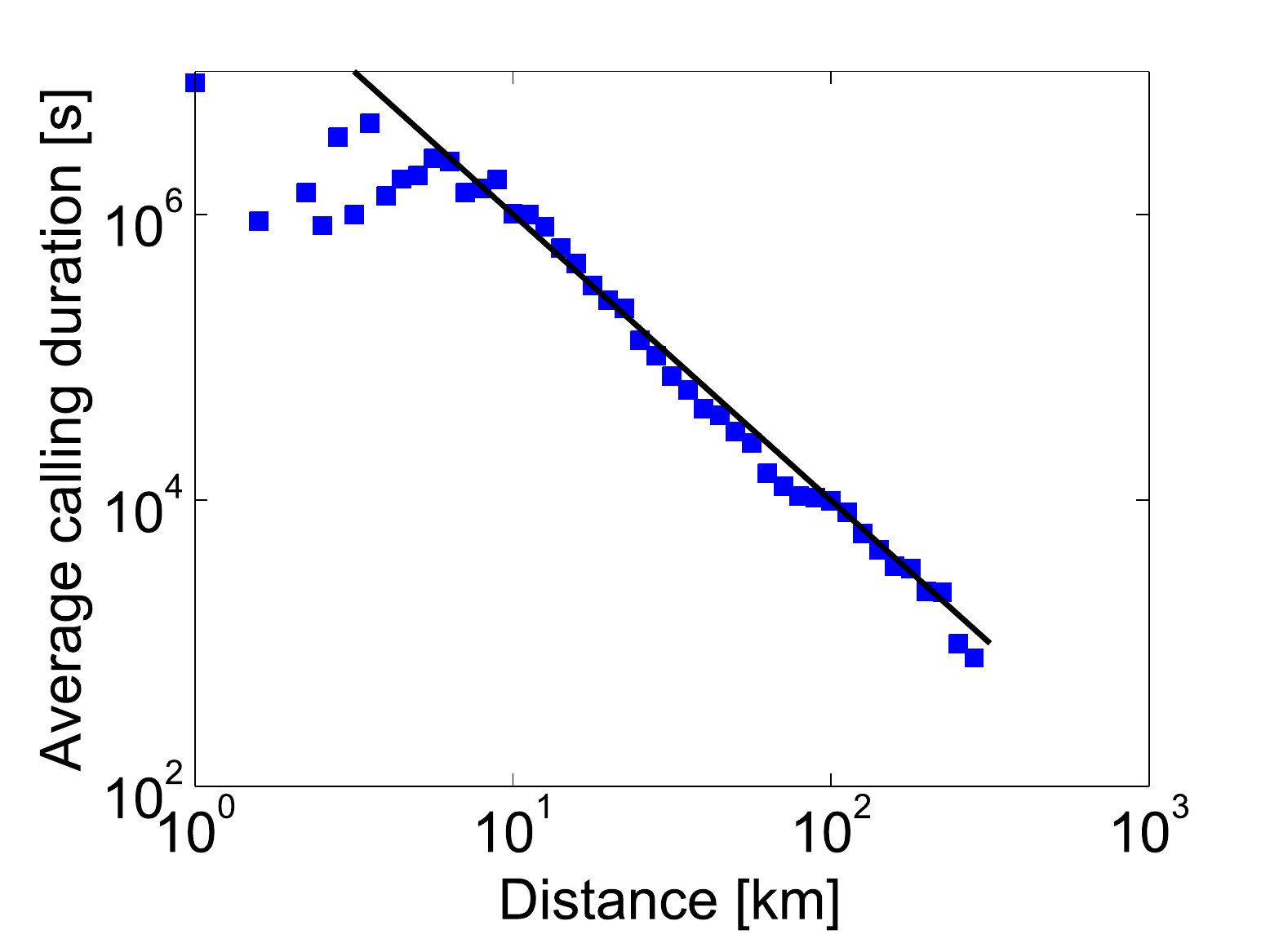}}
	\end{center}
	\caption{(a) Relation between the outgoing intensity $L_{A*}$ (blue), the incoming intensity $L_{*A}$ (red) and the city population size $P_A$. (b) Dependence of the average communication intensity between pairs of cities and the average distance separating them. The black line shows a $\frac{1}{d^2}$ decrease.}
	\label{fig:relationships}
\end{figure}

\begin{figure}[h!]
	\centering
		\includegraphics[scale=0.5]{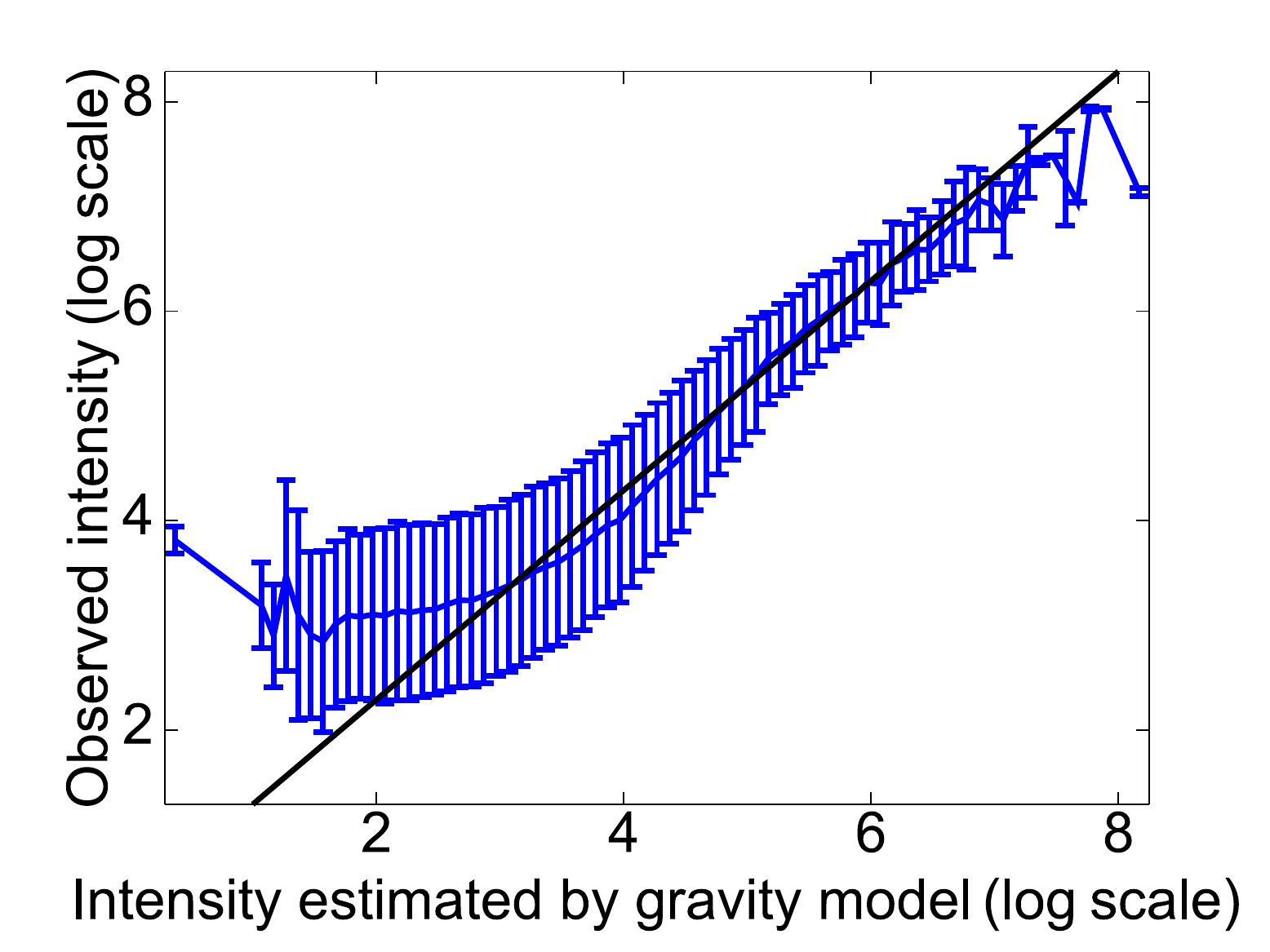}
	\caption{Communication intensity between pairs of cities versus the ratio $\frac{P_AP_B}{d_{AB}^{2}}$. The black line shows the gravitational law.}
	\label{fig:Gravity}
\end{figure}

\begin{figure}[h!]
	\centering
		\includegraphics[scale=0.5]{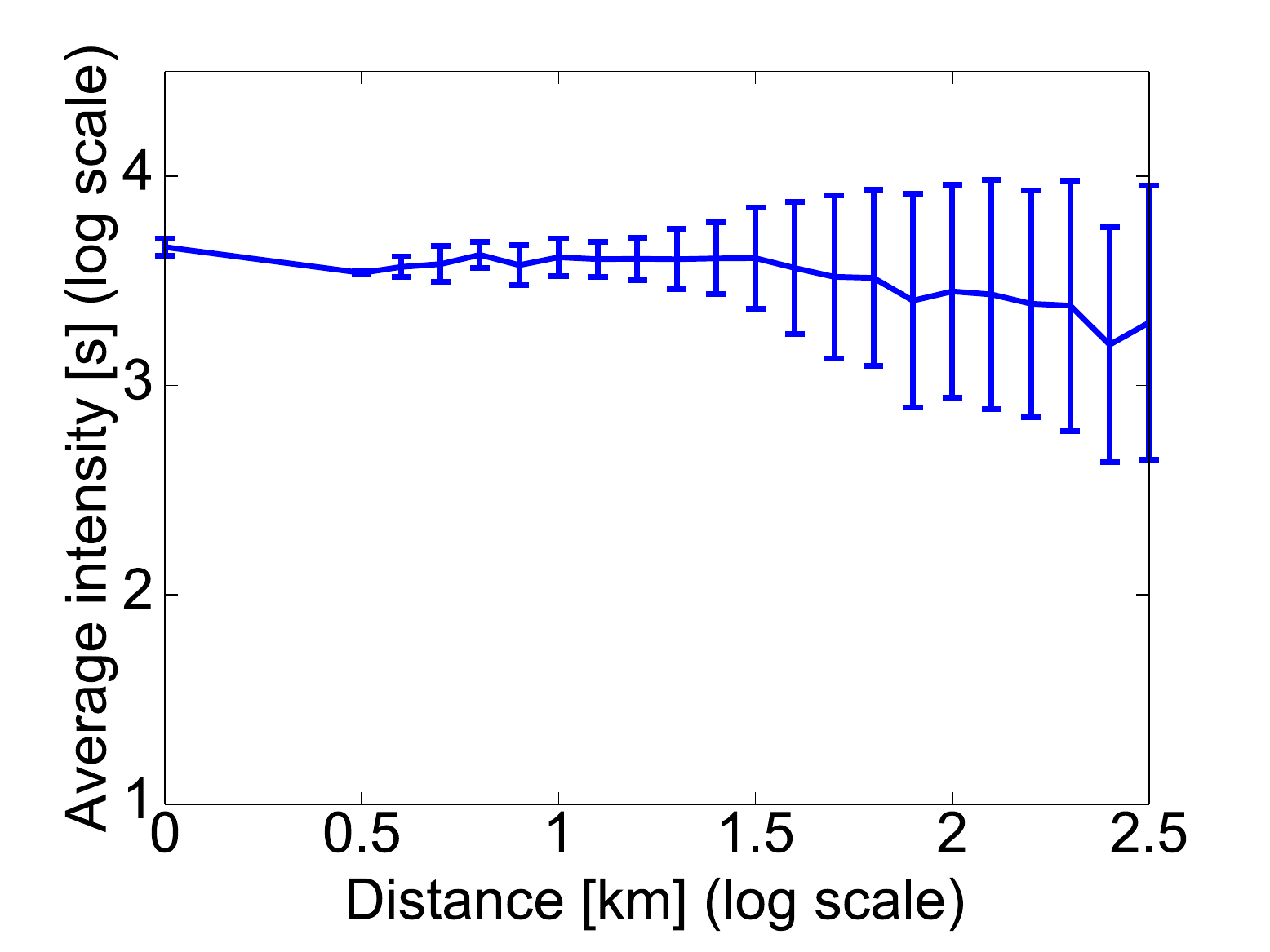}
	\caption{Average intensity of communication between pairs of nodes, if they make a link, versus the distance separating them.}
	\label{fig:TotalTimeCallPeople_distance}
\end{figure}

Many studies have been made on human-to-human communications but few analyses are available on how these communications, once aggregated at the city level, are reliant on the properties of that city. In the following, we model the communication intensity between cities as a function of the population sizes and of the distance between them.\\
First, we analyze how communication flowing into and out of cities, scale with population size. For doing this, for each of the 571 cities we compare the total incoming ($L_{*A}$) and outgoing ($L_{A*}$) communication intensities, as defined below, to the population sizes of these cities.
\begin{equation*}
L_{*A} = \sum_{i \ \notin A,j\in A}l_{ij},\qquad L_{A*} = \sum_{i\in A, j\notin A}l_{ij}.
\label{eq:calls}
\end{equation*}
As shown on Figure \ref{fig:relationships} (a), both incoming and outgoing inter-city communication intensities scale linearly with city size ($L_{A*}, L_{*A} = kP_A^{\beta}$, $\beta=$ 0.96, confidence interval: [0.93 0.99], $R^2=0.87$).
Also, incoming and outgoing communications are strongly symmetric ($L_{A*}\approx L_{*A},\quad\forall A$), that is, calls in one direction  always find a match in the opposite direction.\\
Another parameter that influences communication intensity between cities is distance. It seems reasonable to expect that the intensity of communication between two cities will decay when distance increases and other parameters are kept unchanged. This idea is supported by several studies that suggest gravity-like models for car traffic \cite{jung2008gmk}, trade \cite{bhattacharya2008itn} or economic activity \cite{tinbergen1962awt}. In all these cases, the intensity at distance $d$ is proportional to  $\frac{1}{d^{2}}$. For the case of communication, a similar model has been presented in \cite{zipf1949hba}, but with an intensity evolving like $\frac{1}{d}$, though the authors acknowledge that their model does not seem to fit well with the data.\\
To test the influence of distance on the Belgian network, we measure it as the distance between the centroids of each cities area. The communication intensity between two cities is then compared with the distance between them, where a power law decrease, with an exponent close to $-2$ is shown (see Fig. \ref{fig:relationships} (b)). The power law fits very well for inter-city distances larger than 10 $km$.\\
This result suggests that the communication between cities is ruled by the following gravity model, which is symmetric, scales linearly with the population sizes and decreases with ${d^2}$:
\begin{equation*}
L_{AB} = K\frac{P_AP_B}{d_{AB}^{2}},
\end{equation*}
there, the scaling constant $K$ is the gravity constant for a timespan of 6 months of calling activity.\\
To ensure the validity of our results, we plotted the estimated intensity given by the gravity model versus the observed intensity. As shown on Figure \ref{fig:Gravity}, the results match particularly well for pairs of cities $A$ and $B$ that have a large estimated intensity.
The gravitational constant $K$ can be estimated with a simple best fit of the data. If, over the 6 months of data, the intensity of communication is constant, we obtain the general gravity constant $G=1.07\cdot10^{-2}\frac{s}{day}$. This constant enables us to estimate the intensity of communication between any pair of Belgian cities, based on population, distance and duration of the considered period.
Let us finally observe that this gravity model is consistent with the results presented in \cite{lambiotte2008gdm} that described the probability of connection between customers based on their distance. One can check that the intensity of communication between two customers that make a link does not vary much with the distance between them (see Figure \ref{fig:TotalTimeCallPeople_distance}), so, the distance decay observed in Figure \ref{fig:relationships} (b), does not result from a weaker intensity of communication between customers, but from a smaller fraction of customers communicating with each other.\\
The  gravity model for inter-city communication intensity is analogous to other models of economic and transportation networks, but has to be considered cautiously as it might be biased by the nature of the data.
First of all, Belgium is a small country with a specific demography and two main language communities.
Secondly, we note that our study relies on the definition of census areas, as defined by the Belgian National Institute of Statistics. It is certainly possible that changes to this definition might have non-trivial implications for the shape of the network.
Finally, we only consider the customers of one operator for whom the billing address zip code is available, which means that these customers have a contractual plan.
However, since the observed results suggest clear behavioral influences, it seems unlikely that these results reflect biased or corrupt data. Also, the lognormal shape of city sizes and edge intensities, as well as the power law behavior of calls, is consistent with studies made in other countries using different types of networks.

This work is exploratory, but raises various interesting questions on how people organize their social network depending on the city they live in, as well as, more generally, on the influence of geography on social networks. More significantly, the analysis of social networks, as captured by digital networks like phone calls and email, when overlaid on physical space, could help to improve our understanding of the structure of cities. Or how do they grow into places, where, as Lewis Mumford \cite{mumford1938ch} so eloquently stated: ``the diffused rays of many separate beams of life fall into focus, with gains in both social effectiveness and significance''?

\medskip
\noindent
{\bf Acknowledgements}

{The authors would like to thank Jon Reades and Christine Outram for their help. Gautier Krings acknowledges support from the Concerted Research Action (ARC) ``Large Graphs and Networks'' from the ``Direction de la recherche scientifique - Communaut\'e fran\c{c}aise de Belgique''. The scientific responsibility rests with its authors.}
\section*{References}
\bibliography{CitiesGravity}
\end{document}